\documentclass[fleqn,usenatbib]{mnras}

\usepackage{newtxtext,newtxmath}

\usepackage[T1]{fontenc}
\usepackage{ae,aecompl}

\usepackage{graphicx}	
\usepackage{amsmath}	
\usepackage{amssymb}	
\usepackage{psfig}
\usepackage{xspace}
\usepackage{longtable}
\usepackage{multirow}
\usepackage{natbib}
\graphicspath{{figs/}}
\usepackage{gensymb}
\usepackage{array}

\newcommand{\sref}[1]{Section~\ref{#1}}
\newcommand{\aref}[1]{Appendix~\ref{#1}}
\newcommand{\fref}[1]{Figure~\ref{#1}}

\newcommand{\tref}[1]{Table~\ref{#1}}

\newcolumntype{C}[1]{>{\centering\arraybackslash}m{#1}}
\newcolumntype{L}[1]{>{\arraybackslash}m{#1}}

\title[HSC J0904--0102: A Quadruply-lensed LBG]{Discovery of an unusually compact lensed Lyman Break Galaxy from the Hyper Suprime-Cam Survey} 
\author[A. T. Jaelani et al.]{Anton T. Jaelani,$^{1,2,3}$\thanks{Contact e-mail: \href{anton@phys.kindai.ac.jp}{ anton@phys.kindai.ac.jp}}
Anupreeta More,$^{4,5}$ 
Alessandro Sonnenfeld,$^{4,6}$
Masamune Oguri,$^{4,7,8}$  \newauthor
Cristian E. Rusu,$^{9}$
Kenneth C. Wong,$^{4}$ 
James H. H. Chan,$^{10}$ 
Sherry H. Suyu,$^{11,12,13}$ \newauthor
Issha Kayo,$^{14}$
Chien-Hsiu Lee,$^{15}$
Kaiki T. Inoue$^{1}$ 
\\
\\
$^1$Department of Physics, Kindai University, 3-4-1 Kowakae, Higashi-Osaka, Osaka 577-8502, Japan\\
$^2$Astronomical Institute, Tohoku University, 6-3 Aramaki, Aoba-ku, Sendai 980-8578, Japan\\
$^{3}$Astronomy Study Program and Bosscha Observatory, FMIPA, Institut Teknologi Bandung, Jl. Ganesha 10, Bandung 40132, Indonesia\\
$^{4}$Kavli Institute for the Physics and Mathematics of the Universe (IPMU), 5-1-5 Kashiwanoha, Kashiwa-shi, Chiba 277-8583, Japan\\
$^{5}$The Inter-University Centre for Astronomy and Astrophysics, Post Bag 4, Ganeshkhind, Pune, 411007, India\\
$^{6}$Leiden Observatory, Leiden University, Niels Bohrweg 2, 2333 CA Leiden, Netherlands\\
$^{7}$Department of Physics, The University of Tokyo, 7-3-1 Hongo, Bunkyo-ku, Tokyo 113-0033, Japan\\
$^{8}$Research Center for the Early Universe, The University of Tokyo, 7-3-1 Hongo, Bunkyo-ku, Tokyo 113-0033, Japan\\
$^{9}$Subaru Telescope, National Astronomical Observatory of Japan, 2-21-1 Osawa, Mitaka, Tokyo 181-0015, Japan\\
$^{10}$Laboratoire d'Astrophysique, Ecole Polytechnique F\'{e}d\'{e}rale de Lausanne (EPFL), Observatoire de Sauverny, CH-1290 Versoix, Switzerland\\
$^{11}$Max-Planck-Institut f\"{u}r Astrophysik, Karl-Schwarzschild-Stra{\ss}e 1, 85748 Garching, Germany\\
$^{12}$Institute of Astronomy and Astrophysics, Academia Sinica (ASIAA), 11F of ASMAB, No. 1, Section 4, Roosevelt Road, Taipei 10617, Taiwan\\
$^{13}$Physik-Department, Technische Universit\"{a}t M\"{u}nchen, James-Franck-Stra{\ss}e 1, 85748 Garching, Germany\\
$^{14}$Department of Liberal Arts, Tokyo University of Technology, Ota-ku, Tokyo 144-8650, Japan\\
$^{15}$National Optical Astronomy Observatory (NOAO) 950 N. Cherry Avenue, Tucson, AZ 85719, USA}
\date{Accepted XXX. Received YYY; in original form ZZZ}

\pubyear{2020}

\begin{document}
\label{firstpage}
\pagerange{\pageref{firstpage}--\pageref{lastpage}}
\maketitle

\begin{abstract}
We report the serendipitous discovery of HSC J0904--0102, a quadruply-lensed Lyman break galaxy (LBG) in the Survey of Gravitationally-lensed Objects in Hyper Suprime-Cam Imaging (SuGOHI). Owing to its point-like appearance, the source was thought to be a lensed active galactic nucleus. We obtained follow-up spectroscopic data with the Gemini Multi-Object Spectrographs on the Gemini South Telescope, which confirmed this to be a lens system. The deflecting foreground galaxy is a typical early-type galaxy at a high redshift of $z_{\ell} = 0.957$ with stellar velocity dispersion $\sigma_v=259\pm56$ km~s$^{-1}$. The lensed source is identified as an LBG at $z_{\rm s} = 3.403$, based on the sharp drop bluewards of Ly$\alpha$ and other absorption features. A simple lens mass model for the system, assuming a singular isothermal ellipsoid, yields an Einstein radius of $\theta_{\rm Ein} = 1. 23\arcsec$ and a total mass within the Einstein radius of $M_{\rm Ein} = (5.55\pm 0.24) \times 10^{11}M_{\odot}$ corresponding to a velocity dispersion of $\sigma_{\rm SIE}= 283\pm 3$ km~s$^{-1}$, which is in good agreement with the value derived spectroscopically. The most isolated lensed LBG image has a magnification of $\sim 6.5$. In comparison with other lensed LBGs and typical $z\sim4$ LBG populations, HSC J0904--0102 is unusually compact, an outlier at $>2\sigma$ confidence. Together with a previously discovered SuGOHI lens, HSC J1152+0047, that is similarly compact, we believe that the HSC Survey is extending LBG studies down to smaller galaxy sizes. 
\end{abstract}



\section{Introduction} \label{sec:intro}
Strong gravitational lensing has proven to be a powerful tool for studying the mass distributions of galaxies and clusters, cosmology, and the properties of high redshift sources such as galaxies or quasars. Strongly-lensed sources allow detailed follow-up studies at a fraction of the telescope time that would be necessary for unlensed sources \citep[e.g.,][]{Pettini+02,Stark+07,Quider+09,Quider+10,Richard+08,Richard+11}.

\begin{figure*}
\begin{center}
\includegraphics[width=\textwidth]{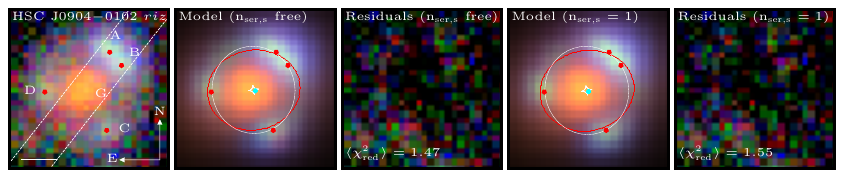}\\
\caption{\label{fig:lensmod} {\bf Panel 1 (left-most) :} HSC $riz$ composite image showing the positions of the quadruply-lensed images (A, B, C, D) and the early-type lens galaxy (G). The locations of peak surface brightness of the lensed images are indicated by red dots. White dashed lines indicate the slit position and orientation used in the spectroscopic observations. North is up and East is left. The white solid bar at the bottom left of the panel shows a scale of 1\arcsec. {\bf Panel 2 :} Model $riz$ composite image generated from lens mass modelling of the lens system, where the source is assumed to be an extended galaxy following a S\'{e}rsic profile (with free S\'{e}rsic index). The best-fit S\'{e}rsic index comes out to be $n_{\rm ser, s}\sim 3$. This panel also shows the unlensed source position (cyan dot). The red contours show the caustics in the source plane and white contours mark the corresponding critical curves in the image plane indicating regions of extremely high magnification. {\bf Panel 3 :} The residual composite image obtained by subtracting the model from the data. {\bf Panel 4 :} Same as the second panel, but for a fixed S\'{e}rsic index, $n_{\rm ser, s}=1$. {\bf Panel 5 (right-most)} : Same as the third panel, but for a fixed S\'{e}rsic index, $n_{\rm ser, s}=1$.}
\end{center}
\end{figure*}

The observed properties of Lyman Break Galaxies (LBGs) provide important constraints on galaxy formation scenarios. In star-forming galaxies at $z\sim 3$, LBG spectra show a "break" around the Lyman limit (912 \AA) in the spectral energy distribution (SED). However, LBGs are generally faint, which makes it difficult to obtain the spectral properties of individual LBGs, so many hundreds are needed to examine their general properties, e.g. the redshift distribution \citep{Adelberger+98,Adelberger+03,Steidel+98}, the rest-frame UV spectroscopic properties \citep{Shapley+03}, the slope of the UV luminosity function \citep{Reddy+09}, and morphology and size evolution \citep[][henceforth, S15]{Shibuya+15}.

The additional magnification provided by strong gravitational lensing enables more detailed investigations of the properties of LBGs. Several lensed LBGs have now been found and studied, e.g. the interstellar medium of the first lensed LBG, MS 1512-cB58 \citep[cB58,][]{Yee+96,Seitz+98,Pettini+00,Pettini+02}, which gives a hint about when most of the metal enrichment occurred. Further studies of cB58 with Spitzer by \citet{Siana+09} have reported that the UV-inferred star-formation rate is lower than that measured from the IR by a factor of 3-5. \citet{Baker+04} found the first direct evidence of the existence of a sizeable cold gas reservoir in an LBG by using observed CO emission from cB58. Similar studies of the "Cosmic Eye" have been carried out \citep{Smail+07,Coppin+07}. Arc-like images of LBGs, lensed by massive galaxy clusters, have been discovered, e.g. A2218--384 \citep{Ebbels+96}, 1E0657-56 \citep{Mehlert+01}, the Sextet Arcs \citep{Frye+07}, SGAS J122651.3+215220 and SGAS J152745.1+065219 \citep{Koester+10}. However, only few additional strongly-lensed LBGs have been lensed by a single massive galaxy, e.g. the Einstein Ring \citep{Cabanac+05}, the Einstein cross \citep{Bolton+06}, the Cosmic Horseshoe \citep{Belokurov+07}, and the 8 o'clock arc \citep{Allam+07}. Studies of current and future lensed LBGs samples will give us better insight when drawing general conclusions for certain properties of the LBG population as a whole. Adding even one lensed LBG would be of tremendous value.

In this paper, we report the discovery and spectroscopic confirmation of a quadruply-lensed LBG from the Hyper Suprime-Cam Subaru (HSC) Survey \citep{Aihara+18}. Our paper is organized as follows. In \sref{sec:data}, we describe the discovery of HSC J0904--0102. We present the spectroscopic follow-up observations on this system, and data reduction procedure in \sref{sec:spec}. In \sref{sec:imdata_ana}, we describe the HSC imaging data analysis and the redshift determinations for the lens and the source. In \sref{sec:lesmodel}, we describe the lens mass modelling of the system. We calculate and discuss the properties of the delensed source in \sref{sec:discussion}. We present our conclusions in \sref{sec:conclusion}. Throughout the paper, we use $\Omega_{\rm m} = 0.27$, $\Omega_{\Lambda}= 0.73$ and  $H_0 = 73$ km s$^{-1}$Mpc$^{-1}$. All quoted magnitudes are on the AB system, all position angles are measured East of North, and all uncertainties are $1\sigma$ and are assumed to be Gaussian.


\section{Discovery of HSC J0904--0102} \label{sec:data}
The HSC Survey is an ongoing imaging survey, expected to cover about 1,400 deg$^2$ in five bands ($g, r, i, z$ and $y$) down to $r\sim 26$ with the Hyper Suprime-Cam \citep{Miyazaki+18,Komiyama+18, Kawanomoto+18, Furusawa+18}, a wide-field (1.7-degree diameter) optical camera installed on the 8.2m Subaru Telescope.  The data  are processed with hscPipe, which is derived from the Large Synoptic Survey Telescope pipeline \citep{Axelrod+10,Juric+17,Ivezic+08,Ivezic+19}. We use photometric data from the HSC Wide S17A internal data release of the HSC survey for our analysis. The seeing in the HSC-$g, r, i, z$ and $y$ images is 0.62\arcsec, 0.76\arcsec, 0.49\arcsec, 0.57\arcsec ~and 0.59\arcsec, respectively.

HSC J0904--0102 was discovered serendipitously during the visual inspection of galaxy clusters selected from the HSC Survey \citep[CAMIRA clusters; for details see][]{Oguri+18} over $\sim$ 232 deg$^2$ of the HSC Wide S16A internal data release. This lens system is comprised of a lens galaxy (G) and four lensed images (A, B, C, D) of the LBG (see left panel of \fref{fig:lensmod}). HSC J0904--0102 is thus part of the sample of lenses discovered from HSC, namely, the Survey of Gravitationally-lensed Objects in HSC Imaging \citep[SuGOHI,][]{Sonnenfeld+18,Sonnenfeld+19,Wong+18,Chan+19,Jaelani+20}.

\begin{figure*}
\begin{center}
\includegraphics[width=\textwidth]{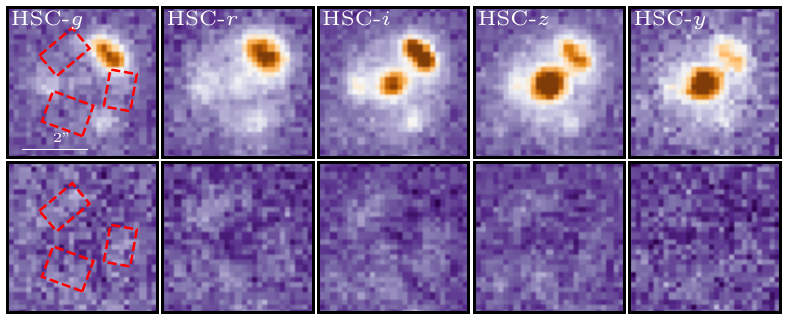}\\
\caption{\label{fig:galfit}{\sc Galfit} modelling results in HSC-$grizy$ images using a S\'{e}rsic source model. Top panels show the HSC data, whereas the bottom panels show {\sc Galfit} model-subtracted residual images in the respective bands. Images are $\sim 5\arcsec$ on the side. The white bar shows a scale of 2\arcsec. Red boxes show the location of very faint extended bluish emission likely arising from the lensed source, consistent with the bluish color of the source.}
\end{center}
\end{figure*}

\begin{table*}
\centering
\caption{\label{tab:lensphotom} Coordinates, photometric redshifts (uncertainties), and magnitudes (uncertainties) in the HSC-$grizy$ data from fitting S\'{e}rsic profiles using {\sc Galfit}.}
\begin{tabular}{C{2cm}C{2cm}cC{2cm}ccccc}
\hline
\hline
RA & Decl. & Name & $z_{\rm  phot}$ & $g$ & $r$ & $i$ & $z$ & $y$\\
(J2000.0) & (J2000.0) &  & err & err & err & err & err & err\\
\hline
09:04:29.75 & $-$01:02:28.26 & G & 1.01 (0.05) & 25.13 (0.11) & 23.12 (0.12) & 21.91 (0.11) & 20.88 (0.06) & 20.59 (0.04)\\
09:04:29.71 & $-$01:02:27.14 & A & 3.26 (0.09) & 24.13 (0.04) & 23.19 (0.03) & 23.01 (0.03) & 22.94 (0.05) & 22.81 (0.09)\\
09:04:29.68 & $-$01:02:27.53 & B & 3.44 (0.05) & 23.91 (0.04) & 23.02 (0.03) & 22.92 (0.04) & 22.87 (0.05) & 22.79 (0.09)\\
09:04:29.71 & $-$01:02:29.47 & C & 3.50 (0.04) & 24.88 (0.09) & 24.03 (0.05) & 23.95 (0.07) & 23.84 (0.11) & 23.61 (0.13)\\
09:04:29.82 & $-$01:02:28.34 & D & 3.49 (0.07) & 25.12 (0.10) & 24.28 (0.06) & 24.08 (0.07) & 24.02 (0.13) & 23.97 (0.13)\\
\hline
 Total LBG &  &  &  & 22.89 (0.03) & 22.00 (0.02) & 21.86 (0.02) & 21.79 (0.04) & 21.68 (0.05)\\
\hline
\end{tabular}
\end{table*}

\section{Spectroscopic Follow-up Observations} \label{sec:spec}
In order to confirm the lensing effect and study the nature of the source galaxy, we performed follow-up spectroscopic observations of HSC J0904--0102 on 2017 November 18 (GS-2017B-FT-5, PI: A. T. Jaelani) with the Gemini Multi-Object Spectrographs (GMOS) on the Gemini South Telescope via the Fast Turnaround program \citep{Mason+14}. The seeing was around 1.0\arcsec$-$1.4\arcsec. GMOS has a wavelength coverage of $\sim$ 4700 \AA~to 9400 \AA \ at a spectral resolution of $R=1918$ using the long-slit mode with the R400-G5305 grating and GG455\_G0305 blocking filter. A slit of width 1\arcsec\ was placed along the merging pair of images and the lens galaxy with a position angle P.A. = $-29\degree$ (see dashed lines in \fref{fig:lensmod}). The total exposure time was 40 minutes. Based on our photometric redshift estimates (see details in \sref{sec:imdata_ana}), we set spatial and spectral dithering with two different central wavelengths (7000 \AA~and 7100 \AA)  to avoid any important lines falling on the gaps between the detectors. The data were binned 2 $\times$ 2, giving a spectral dispersion of 1.48 \AA\ pixel$^{-1}$. 

We used the Image Reduction and Analysis Facility ({\sc IRAF})\footnote{IRAF is distributed by the National Optical Astronomy Observatory, which is operated by the Association of Universities for Research in Astronomy (AURA) under a cooperative agreement with the National Science Foundation.} v2.16 \citep{Tody+86,Tody+93} with the {\sc gmos} package v1.13 \citep{giraf+16} and {\sc Python} tasks ({\sc PyRAF}\footnote{PyRAF is a product of the Space Telescope Science Institute, which is operated by AURA for NASA.}) to reduce the spectra. In general, we followed standard procedure in which the spectra were bias subtracted, flat fielded and sky subtracted using the tasks \texttt{gsflat} and \texttt{gsreduce}. We used the CuAr lamp and a spectroscopic standard star to calibrate the wavelength and flux, respectively (using the tasks \texttt{gswavelength} and \texttt{gstransform} for transformation), then coadded the dithered frames. By using \texttt{gsextract} in an iterative mode, we selected the optimal size and position to extract the lens galaxy and background source spectra. The one-dimensional (1D) spectra of the lens galaxy and lensed images were smoothed with a box 1D convolution kernel of 3.7 \AA. 

\section{Data analysis} \label{sec:imdata_ana}
\subsection{Imaging}
We modelled and analysed the multi-band HSC imaging data for HSC J0904--0102 using {\sc Galfit} \citep{Peng+02} to measure the positions and magnitudes of the lens galaxy and the lensed images. We used the {\sc PSF  Picker} tool to generate PSF models in all of the bands ($g$, $r$, $i$, $z$ and $y$) of the HSC survey for the Wide Field Survey described in \citet{Aihara+18b,Aihara+19}. We find that the lens galaxy in $i$-band is well fit by a model with fixed S\'{e}rsic index $n_{\rm ser, \ell}=4$, ellipticity $e_{\rm \ell}=0.39$, position angle (P.A.) $\theta_{e_{\rm \ell}}= -78\degree$, and effective radius (half-light radius measured along the major axis), $r_{e, {\rm \ell}}=0.71\arcsec$ corresponding to a physical size of $R_{e, {\rm \ell}}\sim 5.49$ kpc at the lens redshift. While the lens galaxy was modelled with a single S\'{e}rsic profile, we tested different models for the lensed images: a PSF model, assuming that the lensed source is point-like, and a S\'{e}rsic  profile, assuming it is extended.

\begin{figure*}
\begin{center}
\includegraphics[width=\textwidth]{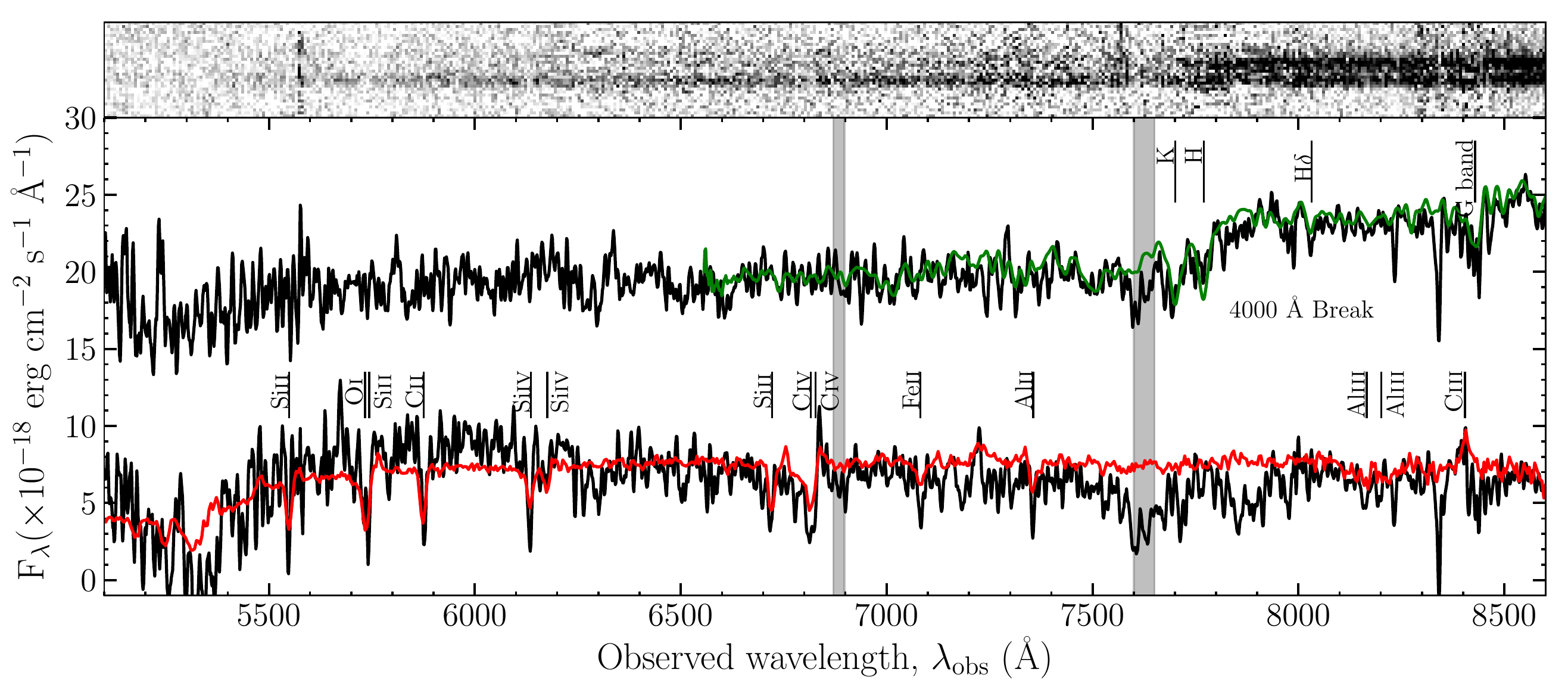}\\
\caption{\label{fig:spectra}Top panel: 2D spectrum of the lens (upper trace) and lensed images (lower trace) from Gemini GMOS spectroscopic follow-up observations. Bottom panel: 1D spectrum of the lens (upper curve) and lensed images (lower curve) superposed with a composite early-type galaxy spectrum \citep{Dobos+12} in green and a composite LBG spectrum \citep{Shapley+03} in red for comparison, respectively, shifted by the measured redshifts. Absorption lines are labelled according to the redshifts of the lens galaxy, $z_{\ell}= 0.957$, and the source galaxy, $z_{\rm s}= 3.403$.}
\end{center}
\end{figure*}

We first modelled the $i$-band data and used the best-fit model as an initial model when fitting in the rest of the bands. In \fref{fig:galfit}, we show all of the multi-band data from HSC and the model-subtracted residual images for the respective bands using a S\'{e}rsic source model. The extended ring-like emission of the underlying host galaxy is revealed in some of the residuals. In general, we found that both choices of source models fit the data equally well, suggesting that the source is highly compact. Magnitudes determined from a single S\'{e}rsic model of the lens galaxy and four separate S\'{e}rsic models of the lensed images are given in \tref{tab:lensphotom}.

Prior to obtaining spectroscopic follow-up data, we also estimated photometric redshifts for the lens galaxy and the lensed images using the publicly available spectral energy distribution (SED) fitting code {\sc LePhare} \citep{Arnouts+99,Ilbert+06}. We used galaxy templates from \cite{Bruzual+03}, subsequently modified by {\sc LePhare} with variable extinction laws \citep{Calzetti+94} and addition of emission lines following  classical recipes that relate galaxy star-formation rates and luminosities in the UV continuum, recombination lines, and forbidden lines \citep{Kennicutt+98}. The source redshift seems to be broadly consistent with the expected redshift $z\sim3$, from the $g-r$ vs. $r-i$ colors \cite[e.g. Figure 4 of][]{Ono+18}. The redshifts thus estimated are given in \tref{tab:lensphotom}.


\subsection{Spectroscopy}
We show the reduced spectra of the lens and the source galaxy in \fref{fig:spectra}. For comparison, we also show composite spectra for early-type galaxies and LBGs from \cite{Dobos+12} and \cite{Shapley+03}, respectively.

We assigned the redshifts of the lens and the source galaxies by identifying a set of lines at a common redshift, fitting a Gaussian profile to each line in order to determine their central wavelength, and taking the mean redshift of the entire set of lines. We measured the redshift of the lens galaxy to be $z_{\ell}=0.957\pm0.003$ based on a strong continuum break around 7800 \AA \ and the characteristic K and H lines, H$\delta$ line and G band present in early-type galaxies. From these lines, we measured a central velocity dispersion $\sigma_v=259\pm56$ km~s$^{-1}$ using the Penalized Pixel Fitting (\textsc{pPXF}) code of \cite{Cappellari+17} using stellar templates from 100 stars (F, G, K, M) selected from the MILES stellar template library of \cite{Sanchez+06}. The mass of the lens galaxy estimated from the measured velocity dispersion $\sigma_v$ and the effective radius $r_e$ using Equation 2 from \citet{Wolf+10} is $M_{\rm {gal}}=(3.4\pm1.5)\times10^{11}M_{\odot}$.

Finally, the source galaxy is confirmed to be $z_{\rm s}=3.403\pm0.001$ from the identification of several absorption features: Si{\sc ii} $\lambda\lambda\lambda(1260.4, 1304.4, 1526.7)$ \AA, O{\sc i} $\lambda1302.2$ \AA, C{\sc ii} $\lambda1334.5$ \AA, Si{\sc iv} $\lambda\lambda(1393.8, 1402.8)$ \AA, C{\sc iv} $\lambda\lambda(1538.2, 1553.8)$ \AA, Fe{\sc ii} $\lambda 1608.5$ \AA, Al{\sc ii} $\lambda 1670.8$ \AA, Al{\sc iii} $\lambda\lambda(1854.7, 1862.8)$ \AA\  and C{\sc iii} $\lambda 1908.7$ \AA. The absorption features and a sharp drop in the flux of the Ly$\alpha$ in the source galaxy spectrum are strong evidences of an LBG galaxy \citep{Steidel+04}. 

\section{Lens Modelling} \label{sec:lesmodel}
We carried out lens mass modelling using the publicly available lens modelling software {\sc glafic} \citep{Oguri+10} to infer the properties of the lens galaxy and the source and to see if there is any improvement to the fits with {\sc Galfit}. The posteriors are sampled with a Markov Chain Monte Carlo (MCMC) technique using the  Python  module {\sc emcee} \citep{Foreman+13}. We run a custom code that combines {\sc glafic} and {\sc emcee} for this purpose. We used the information from each pixel of the multi-band HSC images within a 5\arcsec~box centred on the lens galaxy as our input data constraints for the mass model. 

Light from the lens and source galaxy are each modelled with a seven-parameter elliptical S\'{e}rsic profile. In order to subtract the lens galaxy G, we used the best-fitting parameters from {\sc Galfit} as the initial parameters input to {\sc glafic}. For the source galaxy, we let every parameter of the S\'{e}rsic profile vary: magnitude, position ($\rm x_{s}$, $\rm y_{s}$), position angle, ellipticity, effective radius, and S\'{e}rsic index. 

The mass distribution of the lens galaxy is modelled with a singular isothermal ellipsoid (SIE) profile which has five parameters, namely, centroid ($\rm x$ and $\rm y$), mass, ellipticity and position angle. We fixed the centre of mass to that of the centre of the light profile of the lens galaxy and vary the velocity dispersion $\sigma_{\rm SIE}$ (proxy for mass), ellipticity ($e$), and position angle. HSC J0904--0102 system is located $\sim 195 \arcsec$ (corresponding to $\sim 1500$ kpc) away from a cluster of galaxies at $z=0.864$ with a richness of $\sim 10$. In order to account for additional contribution to the lensing effect from the cluster, we run a test where we allowed for external shear, $\gamma$, to be present in our models. We found $\gamma\approx0.0006$ and did not improve the fit significantly. We also checked from the weak lensing analysis that the contribution of the cluster is negligible at the location of HSC J0904--0102. Hence, we excluded the effect of external shear in our final model.

The best-fit lens model is shown in the second panel of \fref{fig:lensmod} (with reduced $\chi^2_{\rm red}=$ 1.25, 1.67, 1.08, 1.66, 1.72 for the $g, r, i, z, y$-bands, respectively) and the inferred best-fit parameters for the lens and the source are reported in \tref{tab:lensmodel}. The number of degrees of freedom (DOF) is the number of pixels minus the number of the lens and the source parameters used in fitting, $N_{\rm DOF}=444$. We found an effective radius of the source $r_{e, \rm s}=0.022\arcsec$, corresponding to $R_{e, \rm s}\sim 0.16$ kpc with S\'{e}rsic index $n_{\rm ser, s}\sim3$. In order to check the robustness of the source size, we also re-ran the MCMC of the lens model assuming fixed S\'{e}rsic index $n_{\rm ser, s}=1$ and found similar results as when allowing a free S\'{e}rsic index. We also found that the source magnitude in $i$-band, $m_{i, \rm s} = 25.98\pm0.12$, for $n_{\rm ser, s}=1$ was consistent within 1$\sigma$ magnitude for the free S\'{e}rsic index model. The uncertainties were determined using MCMC as plotted in \fref{fig:chaincorner}. We also found that images A and C are located at the minima while images B and D are at saddle points.

The best model is used to predict the Einstein radius $\theta_{\rm Ein}$, defined as 

\begin{equation}\label{eq:Einseq}
\theta_{\rm Ein}=4\pi\left( \frac{\sigma_{\rm SIE}}{c}\right)^2\frac{D_{\rm \ell s}}{D_{\rm s}}
\end{equation}
where $D_{\rm s}$ is the angular diameter distance between the observer and the source, $D_{\rm \ell s}$ is the angular diameter distance between the lens and the source, and $c$ is the speed of light. 

\begin{table}
\centering
\caption{\label{tab:lensmodel} {\bf Top:} Best-fit model parameters in $i$-band with $1\sigma$ errors from MCMC. {\bf Bottom:} The measured or inferred lens and source quantities. Source centroid is relative to the lens centroid.}
\begin{tabular}{L{5.3cm}C{2.3cm}}
\hline
\hline
Lens Parameter  & Best-fit \\
\hline
Velocity dispersion, $\sigma_{\rm SIE}$ (km~s$^{-1}$) & $283\pm 3$\\
Ellipticity, $e_{\rm SIE}$ & 0.17$\pm$0.02 \\
Position Angle, $\theta_{e_{\rm SIE}}$ (degree) & $-80\pm2$ \\
\hline
Source Parameter   & \\
\hline
$i$-magnitude, $m_{i, \rm s}$ & $25.98\pm 0.16$\\
Ellipticity, $e_{\rm s}$ & $0.55\pm0.04$\\
Position Angle, $\theta_{e_{\rm s}}$ (degree) &  $-35\pm 10$ \\
Effective radius, $r_{e, \rm s}$ (arcsec) & $0.022\pm0.007$ \\
S\'{e}rsic index, $n_{\rm ser, s}$ & $2.98\pm 1.25$  \\
\hline
\hline
Quantities & Values \\
\hline
$\Delta$x$_{\rm s}$ (arcsec) & $0.04\pm0.01$\\
$\Delta$y$_{\rm s}$ (arcsec) & $-0.03\pm0.01$\\
Spectroscopic lens redshift, $z_{\ell}$ & $0.957\pm0.003$ \\
Spectroscopic source redshift, $z_{\rm s}$ & $3.403\pm0.001$ \\
Einstein Radius, $\theta_{\rm Ein}$ (arcsec) & $1.23\arcsec\pm 0.03\arcsec$ \\
Mass within $\theta_{\rm Ein}$,  $M_{\rm Ein}$ ($M_{\odot}$) & $(5.55\pm0.24)\times 10^{11}$ \\
\hline
\end{tabular}
\end{table}

\section{Discussion} \label{sec:discussion}
In this section, we study and explore the reasons that might have contributed to the detection of a system like HSC J0904--0102 including any biases from strong lensing and the robustness of the inferred compactness, by testing for systematic errors introduced due to lens modelling.

\subsection{Intrinsic source properties}

\begin{figure}
\begin{center}
\setlength{\leftskip}{-0.3cm}
\includegraphics[width=0.49\textwidth]{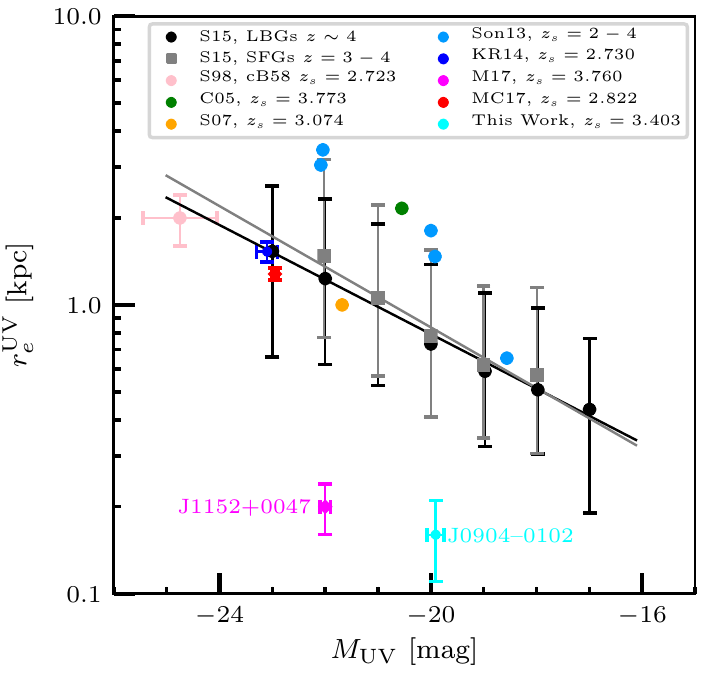}\\
\caption{\label{fig:shibuya15} Effective radius and absolute UV magnitude. HSC J0904--0102 is shown along with other lensed LBG or LAE systems at redshift $z\sim 3$ from \citet[][S98]{Seitz+98}, \citet[][C05]{Cabanac+05}, \citet[][S07]{Smail+07}, \citet[][Son13]{Sonnenfeld+13a,Sonnenfeld+13b}, \citet[][KR14]{Kostrzewa+14}, \citet[][M17]{More+17}, and \citet[][MC17]{Marques+17}. For comparison, we show the size-luminosity relation of LBGs (black line) and star-forming galaxies (SFGs, grey line) from S15. The line denotes their best-fit for the $r_e- M_{\rm UV}$ relation and their data points with error bars indicate the median $r_e$ and the 16th and 84th percentiles for their sample.}
\end{center}
\end{figure}

\begin{figure*}
\begin{center}
\includegraphics[width=\textwidth]{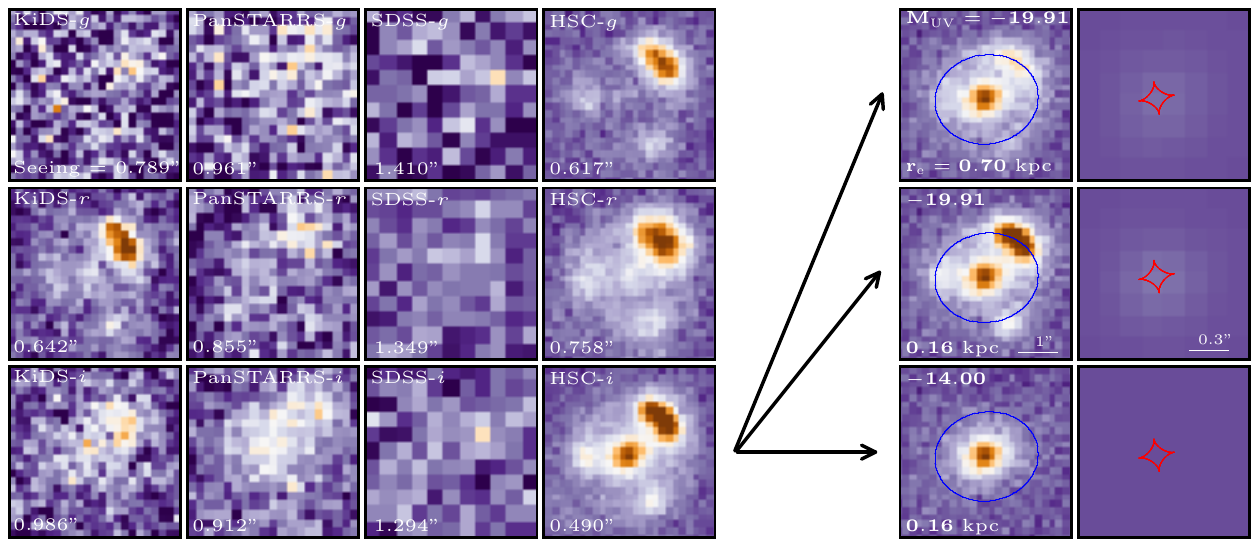}\\
\caption{\label{fig:shibuya15images} Left panels: HSC J0904--0102 images from other overlapping surveys (KiDS, PanSTARRS, and SDSS), along with images from the HSC Survey for $g$, $r$, and $i$-bands. Right panels: Lens model where lens parameters are fixed to the best-fitting parameters, while the source was modelled as a S\'{e}rsic profile with three different conditions: $r_e$ fixed to $\sim 0.7$ kpc (top right panel), best-fitting parameters (middle right panel), and $M_{\rm UV}$ fixed to $\sim -14$ (bottom right panel) based one the $r_e- M_{\rm UV}$ relation as derived by S15 with the other parameters from the best-fit. Blue lines indicate critical curves. True source images for corresponding conditions are shown in right-most panels with caustic lines (red).}
\end{center}
\end{figure*}

Having determined the lensing magnification and the effective radius of the LBG, we can estimate its intrinsic luminosity and size. We calculate the absolute UV (at rest-frame $1700$ \AA) magnitude from the apparent magnitude of the source and redshift $z_s$ using
\begin{equation}\label{eq:absM}
M_{\rm UV}=m+2.5\log (1+z_s)-5\log \left(\frac{D_{\rm L}(z_s)}{10 \rm pc}\right)+(m_{\rm UV}-m)
\end{equation}
where $D_{\rm L}$ is the luminosity distance in the unit parsecs, $m$ is the apparent magnitude, and $(m_{\rm UV}-m)$ is the $k-$correction term between the magnitude at rest-frame UV and the observed magnitude that we use. In order to correct for reddening by dust extinction, we estimated the $k-$correction term by assuming the intrinsic UV continuum slope from Figure 3 in \citet{Sawicki+06}. 

Correcting for the estimated magnification and UV reddening, we determine the absolute UV magnitude using two different $k-$corrections: the $100$ Myr and $10$ Myr-old continuously star-forming template with a dust attenuation of $E(B-V)=0.2$ and $0.0$, respectively. For the $k-$correction with $[100\ \rm Myr,$ $E(B-V)=0.2]$, we determine $M_{\rm 1700}=-19.73\pm 0.16$ corresponding to $L_{1700}\sim 0.34\times 10^{29}$ ergs s$^{-1}$ Hz$^{-1}$. For $k-$correction with $[10\ \rm Myr,$ $ E(B-V)=0.0]$, we determine $M_{\rm 1700}=-19.91\pm 0.16$ corresponding to $L_{1700}\sim 0.40\times 10^{29}$ ergs s$^{-1}$ Hz$^{-1}$. 

Following the same approach as \cite{More+17}, we estimate the true source magnitude and size in a manner complementary to the method described in \sref{sec:lesmodel}. Using the magnification factor of image C (which is the most isolated lensed LBG image), $\mu_{\rm C}\sim 6.5$, we obtain a delensed source magnitude of $m_{i, \rm \ell}=25.98$. This result was consistent within 1$\sigma$ magnitude for the best-fitting source magnitude, $m_{i, \rm \ell} = 25.98\pm 0.16$. The true source size is estimated to be $r_{e, \rm s\ true} = \sqrt{r_{e,\  \rm Galfit}^2/\mu_{\rm C}}=0.02\arcsec\pm 0.01$\arcsec, where $r_{e,\  \rm Galfit}= 0.05\arcsec\pm0.02\arcsec$, corresponds to $R_{e,\rm s\ true}\sim0.14\pm 0.06$ kpc in physical units. We derive an intrinsic star-formation rate of $\sim 5 M_{\odot}$ yr$^{-1}$. HSC J0904--0102 is intrinsically less luminous relative to the typical luminosities of LBGs, i.e., $L_{\rm UV}=1.06\times10^{29}$ ergs s$^{-1}$ Hz$^{-1}$ which corresponds to  $m_{\rm UV} (\simeq 1500-1700$ \AA) = 24.61 \citep{Reddy+09}.

Our estimated size from both {\sc glafic} and {\sc Galfit} suggest a size much smaller than the typical LBGs at $z\sim4$ and star-forming galaxies (SFGs) at $z\sim3-4$ (see \fref{fig:shibuya15} and S15). The SFGs and LBGs in S15 are identified from deep HST surveys (e.g. HUDF, CANDELS and HFF \footnote{Hubble Ultra Deep Field, Cosmic Assembly Near-infrared Deep Extragalactic Legacy Survey, Hubble Frontier Field.}) and are apparently faint sources (optical/UV mag$\sim23-28$ for redshifts spanning $z\sim0-10$). Even though they are observed with HST resolution, these sources are bigger than our lensed LBG. On the other hand, the combined imaging area investigated in S15 is much smaller (a few sq. deg.) compared to the HSC Survey (hundreds of sq. deg. in this work). Thus, it is possible that the work of S15 is missing highly compact LBGs, such as our source, if they come from a rarer population.   

If the true lensing magnification is higher than our currently estimated value by factor $\mu$, then it would mean that the source is intrinsically fainter by $\mu$ and smaller by $\sqrt{\mu}$, which would move the source to rightwards and downwards on \fref{fig:shibuya15}. On the contrary, if the true magnification is lower, then our source would move leftwards and upwards on \fref{fig:shibuya15} while preserving the surface brightness, nearly parallel to the S15 relation.

\begin{table*}
\centering
\caption{\label{tab:mockparam} Mock and best-fit model parameters.}
\begin{tabular}{L{6cm} C{1.6cm} C{1.6cm} C{0.6cm} C{1.6cm} C{1.6cm} C{1.6cm}}
\hline
\hline
Lens Parameter   & {\bf Mock1} & Model1 & & {\bf Mock2} & Model2 (a) & Model2 (b)\\
\hline
Velocity dispersion, $\sigma_{\rm SIE}$ (km~s$^{-1}$) & 320 & 320 $\pm$ 1 & & -- & 324 $\pm$ 2 & --\\
Einstein Radius, $\theta_{\rm Ein}$ (arcsec) & -- & -- & & 1.60 & -- & 1.58 $\pm$ 0.04\\
Ellipticity, $e_{\rm SIE}$ & 0.25 & 0.24 $\pm$ 0.01 & & 0.25 & 0.33 $\pm$ 0.02 & 0.25 $\pm$ 0.03\\
Position Angle, $\theta_{e_{\rm SIE}}$ (degree)  & $-$81 & $-$81 & & 20 & 19.53 $\pm$ 0.55 & 19 $\pm$ 1\\
Slope & 2 & 2 & & 2.3 & 2 & 2.18 $\pm$ 0.17\\
\hline
Source Parameter   & & & & &  &\\
\hline
Magnitude & 25.15 & 25.19 $\pm$ 0.12 & & 24.02 & 24.34 $\pm$ 0.06 & 23.90 $\pm$ 0.11\\
Ellipticity,  $e_{\rm s}$ & 0.40 & 0.23 $\pm$ 0.16 & & 0.4 & 0.55 $\pm$ 0.08 & 0.15 $\pm$ 0.01\\
Position Angle, $\theta_{e_{\rm s}}$ (degree)  & $-$51 & $-$51 & & $-$51 & $-$56 $\pm$ 4 & $-$77 $\pm$ 5\\
Effective radius, $r_{e, \rm s}$ (arcsec) & 0.08 & 0.08 $\pm$ 0.02 & & 0.2 & 0.25 $\pm$ 0.03 & 0.23 $\pm$ 0.01\\
S\'{e}rsic index, $n_{\rm ser, s}$ & 3.50 & 3.23 $\pm$ 1.59 & & 1.5 & 2.82 $\pm$ 0.54 & 0.38 $\pm$ 0.08\\
\hline
\end{tabular}
\end{table*}

In \fref{fig:shibuya15}, we show another quad, lens HSC~J1152+0047 \citep[][henceforth, referred to as J1152]{More+17}, also discovered serendipitously from HSC. The J1152 quad also happens to be unusually compact compared to the rest of the LBGs or Lyman Alpha Emitters (LAEs) at $z\sim4$. Additionally, we compare to other lensed LBGs from the literatur at $z\sim2-4$ for which we could derive the sizes and magnitudes. Their sizes and absolute magnitudes are consistent with the typical LBG population of S15. A few others that could not be added here (due to missing sizes or absolute magnitudes) for a quantitative comparison showed fairly extended arc-like emission, suggesting that the intrinsic sources are unlikely to be highly compact. Overall, we did not find any lensed LBGs in the literature as compact as our source and J1152. 

Similar to J1152, our lens system is also located in the footprint of other surveys such as the KiDS \citep{Kuijken+19}, PanSTARRS \citep{Chambers+16}, and SDSS \citep[DR15][]{Aguado+19} but is barely detectable or identifiable as a lens owing to the shallower depth of those surveys coupled with the faintness of the lens system (see panels on the left in \fref{fig:shibuya15images}). We show how the appearance of lensed images would change in HSC imaging if the source were to be a typical galaxy following the size-magnitude relation of S15. First, we fix the source magnitude to the best-fit model result and estimate the source size (following S15). This makes the source bigger (top-right panels) and hence, harder to detect in HSC imaging owing to the decreased surface brightness. Next, we fix the source size to the best-fit model result and estimate its magnitude (following S15). This makes the source fainter (bottom-right panels) and yet again, impossible to detect in HSC imaging. For reference, we show the best-fit source model (middle-right panels) in \fref{fig:shibuya15images}.

Since S\'{e}rsic index can affect the size and magnitude parameters, we also tested models by varying the S\'{e}rsic index parameter ($n_{\rm ser, s}=1$ or free). We find that the data is not able to constrain the S\'{e}rsic index accurately, but the inferred source size is measured more robustly. Finally, we note that differential magnification due to lensing may create a bias in favor of the detection of more compact sources than their bigger counterparts \citep{Hezaveh+12,Oldham+17}. While this may be important, we believe that lensing bias may not be the primary reason for the detection of our source, as explained above. Instead, it is the unique combination of deep, wide imaging with superior image quality (PSF) including better pixel resolution of HSC data that has allowed us to discover a rare and highly compact LBG.  

\subsection{Systematic uncertainties on the source size due to lens modelling}

We estimate the systematic uncertainties of lens modelling using two mock systems. We used {\sc glafic} to generate mocks. The generated mocks are similar to HSC J0904--0102, that is, a single lens and a single source with redshifts and positions matched to the real lens. First, AM generated a mock lens with an SIE density profile as the lens model. The lens mass, lens ellipticity and the following source parameters were unknown to the modeller (ATJ) - flux, ellipticity, effective radius and S\'{e}rsic index. The rest of the parameters along with the choice of density profile were known to ATJ. The results of the modelling are given in \tref{tab:mockparam}. The best-fit effective radius and flux are consistent with the input values. 

Next, AM generated a mock lens with a power-law density profile and a slope of 2.3 instead of an SIE (slope=2). ATJ modelled the system without any information about the choice of density profile or any of the parameter values. The corresponding best-fit is given in \tref{tab:mockparam} assuming an SIE model first, and as a result the lens parameters are not fit as well. However, the source size is still consistent with the input value, while the intrinsic source flux is somewhat dependent on the choice of the mass density profile. In the next iteration, ATJ had the information that the density profile is not SIE. This improved the fit to the lens model parameters and the source size remained consistent with the input value within the uncertainties, as before. We point out that the source size and fluxes are constrained robustly in the above circumstances.

The source sizes in our mocks bracket the value of $\sim 0.1\arcsec$, which is the size our source would have if it were to follow the Shibuya et al. relation. Since we are able to constrain the "tested" sizes with reasonable accuracy, we are confident that our inferred $r_{e, \rm s}$ is most likely not as large as $0.1\arcsec$.


\section{Conclusion}
\label{sec:conclusion}
We report the discovery and spectroscopic confirmation of a quadruply-lensed LBG at $z=3.403$. The background source is strongly-lensed by a galaxy at a high redshift of $z=0.957$ with a magnification factor for the most isolated lensed LBG image of $6.5$. A simple SIE lens model fits the lens system well and suggests an Einstein radius $\theta_{\rm Ein}=1.23\arcsec\pm0.03\arcsec$ ($R_{\rm Ein}=9.48\pm0.20$ kpc). This implies that the mass($<R_{\rm Ein}$) is $5.55 \times 10^{11}M_{\odot}$. We also infer intrinsic source properties, primarily the source size, which, given the magnification factor, turns out to be 0.14 kpc. In comparison with other LBGs (both lensed and unlensed) from the literature, the HSC J0904--0102 source turns out to be unusually compact (0.16 kpc) and 1$\sigma$ smaller than other LBGs (both lensed and unlensed) from the literatures. Not only is the LBG apparently faint, but it also is intrinsically less luminous (about 0.4$L_*$), corresponding to a star-formation rate of $\sim 5 M_{\odot}$ yr$^{-1}$. Along with J1152 \citep{More+17}, another unusually compact lensed source from the HSC Survey, the discovery of HSC J0904--0102 and based on our rough estimate (see \aref{app:expected}), we expect to discover at least 40 unusually compact lensed system by the end of the HSC Survey. Therefore, it is possible that we are beginning to unearth a source population that could not be studied before.


\section*{Acknowledgements}
We would like to thank Masami Ouchi, Yoshiaki Ono, and Yuichi Higuchi for useful discussions. The authors would like to thank the referee for improving the presentation of the paper. ATJ and KTI are supported by JSPS KAKENHI Grant Number JP17H02868. MO acknowledges support from JSPS KAKENHI Grant Numbers JP15H05892 and JP18K03693. SHS thanks the Max Planck Society for support through the Max Planck Research Group. IK was supported in part by JSPS KAKENHI Grant Number JP15H05896. This work was supported in part by World Premier International Research Center Initiative (WPI Initiative), MEXT, Japan. The Hyper Suprime-Cam (HSC) collaboration includes the astronomical communities of Japan and Taiwan, and Princeton University. The HSC instrumentation and software were developed by the National Astronomical Observatory of Japan (NAOJ), the Kavli Institute for the Physics and Mathematics of the Universe (Kavli IPMU), the University of Tokyo, the High Energy Accelerator Research Organization (KEK), the Academia Sinica Institute for Astronomy and Astrophysics in Taiwan (ASIAA), and Princeton University. Funding was contributed by the FIRST program from Japanese Cabinet Office, the Ministry of Education, Culture, Sports, Science and Technology (MEXT), the Japan Society for the Promotion of Science (JSPS), Japan Science and Technology Agency (JST), the Toray Science Foundation, NAOJ, Kavli IPMU, KEK, ASIAA, and Princeton University. This paper makes use of software developed for the Large Synoptic Survey Telescope. We thank the LSST Project for making their code available as free software at \href{http://dm.lsst.org}{http://dm.lsst.org}. This paper is based [in part] on data collected at the Subaru Telescope and retrieved from the HSC data archive system, which is operated by Subaru Telescope and Astronomy Data Center at National Astronomical Observatory of Japan. Data analysis was in part carried out with the cooperation of Center for Computational Astrophysics, National Astronomical Observatory of Japan. The Pan-STARRS1 Surveys (PS1) and the PS1 public science archive have been made possible through contributions by the Institute for Astronomy, the University of Hawaii, the Pan-STARRS Project Office, the Max-Planck Society and its participating institutes, the Max Planck Institute for Astronomy, Heidelberg and the Max Planck Institute for Extraterrestrial Physics, Garching, The Johns Hopkins University, Durham University, the University of Edinburgh, the Queen's University Belfast, the Harvard-Smithsonian Center for Astrophysics, the Las Cumbres Observatory Global Telescope Network Incorporated, the National Central University of Taiwan, the Space Telescope Science Institute, the National Aeronautics and Space Administration under Grant No. NNX08AR22G issued through the Planetary Science Division of the NASA Science Mission Directorate, the National Science Foundation Grant No. AST-1238877, the University of Maryland, Eotvos Lorand University (ELTE), the Los Alamos National Laboratory, and the Gordon and Betty Moore Foundation. Based on observations obtained at the Gemini Observatory, which is operated by the Association of Universities for Research in Astronomy, Inc., under a cooperative agreement with the NSF on behalf of the Gemini partnership: the National Science Foundation (United States), National Research Council (Canada), CONICYT (Chile), Ministerio de Ciencia, Tecnolog\'{i}a e Innovaci\'{o}n Productiva (Argentina), Minist\'{e}rio da Ci\^{e}ncia, Tecnologia e Inova\c{c}\~{a}o (Brazil), and Korea Astronomy and Space Science Institute (Republic of Korea). The authors wish to recognize and acknowledge the very significant cultural role and reverence that the summit of Mauna Kea has always had within the indigenous Hawaiian community. We are most fortunate to have the opportunity to conduct observations from this mountain. Based on observations made with ESO Telescopes at the La Silla Paranal Observatory under programme IDs 177.A-3016, 177.A-3017, 177.A-3018 and 179.A-2004, and on data products produced by the KiDS consortium. The KiDS production team acknowledges support from: Deutsche Forschungsgemeinschaft, ERC, NOVA and NWO-M grants; Target; the University of Padova, and the University Federico II (Naples).

\bibliographystyle{mnras}
\bibliography{references_papers}

\begin{thebibliography}{}
\makeatletter
\relax
\def\mn@urlcharsother{\let\do\@makeother \do\$\do\&\do\#\do\^\do\_\do\%\do\~}
\def\mn@doi{\begingroup\mn@urlcharsother \@ifnextchar [ {\mn@doi@}
  {\mn@doi@[]}}
\def\mn@doi@[#1]#2{\def\@tempa{#1}\ifx\@tempa\@empty \href
  {http://dx.doi.org/#2} {doi:#2}\else \href {http://dx.doi.org/#2} {#1}\fi
  \endgroup}
\def\mn@eprint#1#2{\mn@eprint@#1:#2::\@nil}
\def\mn@eprint@arXiv#1{\href {http://arxiv.org/abs/#1} {{\tt arXiv:#1}}}
\def\mn@eprint@dblp#1{\href {http://dblp.uni-trier.de/rec/bibtex/#1.xml}
  {dblp:#1}}
\def\mn@eprint@#1:#2:#3:#4\@nil{\def\@tempa {#1}\def\@tempb {#2}\def\@tempc
  {#3}\ifx \@tempc \@empty \let \@tempc \@tempb \let \@tempb \@tempa \fi \ifx
  \@tempb \@empty \def\@tempb {arXiv}\fi \@ifundefined
  {mn@eprint@\@tempb}{\@tempb:\@tempc}{\expandafter \expandafter \csname
  mn@eprint@\@tempb\endcsname \expandafter{\@tempc}}}

\bibitem[\protect\citeauthoryear{{Adelberger}, {Steidel}, {Giavalisco},
  {Dickinson}, {Pettini}  \& {Kellogg}}{{Adelberger}
  et~al.}{1998}]{Adelberger+98}
{Adelberger} K.~L.,  {Steidel} C.~C.,  {Giavalisco} M.,  {Dickinson} M.,
  {Pettini} M.,   {Kellogg} M.,  1998, \mn@doi [\apj] {10.1086/306162}, \href
  {http://adsabs.harvard.edu/abs/1998ApJ...505...18A} {505, 18}

\bibitem[\protect\citeauthoryear{{Adelberger}, {Steidel}, {Shapley}  \&
  {Pettini}}{{Adelberger} et~al.}{2003}]{Adelberger+03}
{Adelberger} K.~L.,  {Steidel} C.~C.,  {Shapley} A.~E.,   {Pettini} M.,  2003,
  \mn@doi [\apj] {10.1086/345660}, \href
  {http://adsabs.harvard.edu/abs/2003ApJ...584...45A} {584, 45}

\bibitem[\protect\citeauthoryear{{Aguado} et~al.,}{{Aguado}
  et~al.}{2019}]{Aguado+19}
{Aguado} D.~S.,  et~al., 2019, \mn@doi [\apjs] {10.3847/1538-4365/aaf651},
  \href {https://ui.adsabs.harvard.edu/abs/2019ApJS..240...23A} {240, 23}

\bibitem[\protect\citeauthoryear{{Aihara} et~al.,}{{Aihara}
  et~al.}{2018a}]{Aihara+18}
{Aihara} H.,  et~al., 2018a, \mn@doi [\pasj] {10.1093/pasj/psx066}, \href
  {http://adsabs.harvard.edu/abs/2018PASJ...70S...4A} {70, S4}

\bibitem[\protect\citeauthoryear{{Aihara} et~al.,}{{Aihara}
  et~al.}{2018b}]{Aihara+18b}
{Aihara} H.,  et~al., 2018b, \mn@doi [\pasj] {10.1093/pasj/psx081}, \href
  {https://ui.adsabs.harvard.edu/abs/2018PASJ...70S...8A} {70, S8}

\bibitem[\protect\citeauthoryear{{Aihara} et~al.,}{{Aihara}
  et~al.}{2019}]{Aihara+19}
{Aihara} H.,  et~al., 2019, arXiv e-prints, \href
  {https://ui.adsabs.harvard.edu/abs/2019arXiv190512221A} {p. arXiv:1905.12221}

\bibitem[\protect\citeauthoryear{{Allam}, {Tucker}, {Lin}, {Diehl}, {Annis},
  {Buckley-Geer}  \& {Frieman}}{{Allam} et~al.}{2007}]{Allam+07}
{Allam} S.~S.,  {Tucker} D.~L.,  {Lin} H.,  {Diehl} H.~T.,  {Annis} J.,
  {Buckley-Geer} E.~J.,   {Frieman} J.~A.,  2007, \mn@doi [\apjl]
  {10.1086/519520}, \href
  {https://ui.adsabs.harvard.edu/abs/2007ApJ...662L..51A} {662, L51}

\bibitem[\protect\citeauthoryear{{Arnouts}, {Cristiani}, {Moscardini},
  {Matarrese}, {Lucchin}, {Fontana}  \& {Giallongo}}{{Arnouts}
  et~al.}{1999}]{Arnouts+99}
{Arnouts} S.,  {Cristiani} S.,  {Moscardini} L.,  {Matarrese} S.,  {Lucchin}
  F.,  {Fontana} A.,   {Giallongo} E.,  1999, \mn@doi [\mnras]
  {10.1046/j.1365-8711.1999.02978.x}, \href
  {http://adsabs.harvard.edu/abs/1999MNRAS.310..540A} {310, 540}

\bibitem[\protect\citeauthoryear{{Axelrod}, {Kantor}, {Lupton}  \&
  {Pierfederici}}{{Axelrod} et~al.}{2010}]{Axelrod+10}
{Axelrod} T.,  {Kantor} J.,  {Lupton} R.~H.,   {Pierfederici} F.,  2010, in
  Software and Cyberinfrastructure for Astronomy. p. 774015,
  \mn@doi{10.1117/12.857297}

\bibitem[\protect\citeauthoryear{{Baker}, {Tacconi}, {Genzel}, {Lehnert}  \&
  {Lutz}}{{Baker} et~al.}{2004}]{Baker+04}
{Baker} A.~J.,  {Tacconi} L.~J.,  {Genzel} R.,  {Lehnert} M.~D.,   {Lutz} D.,
  2004, \mn@doi [\apj] {10.1086/381798}, \href
  {https://ui.adsabs.harvard.edu/abs/2004ApJ...604..125B} {604, 125}

\bibitem[\protect\citeauthoryear{{Belokurov} et~al.,}{{Belokurov}
  et~al.}{2007}]{Belokurov+07}
{Belokurov} V.,  et~al., 2007, \mn@doi [\apjl] {10.1086/524948}, \href
  {http://adsabs.harvard.edu/abs/2007ApJ...671L...9B} {671, L9}

\bibitem[\protect\citeauthoryear{{Bolton}, {Moustakas}, {Stern}, {Burles},
  {Dey}  \& {Spinrad}}{{Bolton} et~al.}{2006}]{Bolton+06}
{Bolton} A.~S.,  {Moustakas} L.~A.,  {Stern} D.,  {Burles} S.,  {Dey} A.,
  {Spinrad} H.,  2006, \mn@doi [\apjl] {10.1086/506446}, \href
  {http://adsabs.harvard.edu/abs/2006ApJ...646L..45B} {646, L45}

\bibitem[\protect\citeauthoryear{{Bruzual} \& {Charlot}}{{Bruzual} \&
  {Charlot}}{2003}]{Bruzual+03}
{Bruzual} G.,  {Charlot} S.,  2003, \mn@doi [\mnras]
  {10.1046/j.1365-8711.2003.06897.x}, \href
  {http://adsabs.harvard.edu/abs/2003MNRAS.344.1000B} {344, 1000}

\bibitem[\protect\citeauthoryear{{Cabanac}, {Valls-Gabaud}, {Jaunsen}, {Lidman}
   \& {Jerjen}}{{Cabanac} et~al.}{2005}]{Cabanac+05}
{Cabanac} R.~A.,  {Valls-Gabaud} D.,  {Jaunsen} A.~O.,  {Lidman} C.,   {Jerjen}
  H.,  2005, \mn@doi [\aap] {10.1051/0004-6361:200500115}, \href
  {http://adsabs.harvard.edu/abs/2005A%26A...436L..21C} {436, L21}

\bibitem[\protect\citeauthoryear{{Calzetti}, {Kinney}  \&
  {Storchi-Bergmann}}{{Calzetti} et~al.}{1994}]{Calzetti+94}
{Calzetti} D.,  {Kinney} A.~L.,   {Storchi-Bergmann} T.,  1994, \mn@doi [\apj]
  {10.1086/174346}, \href {http://adsabs.harvard.edu/abs/1994ApJ...429..582C}
  {429, 582}

\bibitem[\protect\citeauthoryear{{Cappellari}}{{Cappellari}}{2017}]{Cappellari+17}
{Cappellari} M.,  2017, \mn@doi [\mnras] {10.1093/mnras/stw3020}, \href
  {http://adsabs.harvard.edu/abs/2017MNRAS.466..798C} {466, 798}

\bibitem[\protect\citeauthoryear{{Chambers} et~al.,}{{Chambers}
  et~al.}{2016}]{Chambers+16}
{Chambers} K.~C.,  et~al., 2016, arXiv e-prints, \href
  {https://ui.adsabs.harvard.edu/abs/2016arXiv161205560C} {p. arXiv:1612.05560}

\bibitem[\protect\citeauthoryear{{Chan} et~al.,}{{Chan} et~al.}{2019}]{Chan+19}
{Chan} J. H.~H.,  et~al., 2019, arXiv e-prints, \href
  {https://ui.adsabs.harvard.edu/abs/2019arXiv191102587C} {p. arXiv:1911.02587}

\bibitem[\protect\citeauthoryear{{Coppin} et~al.,}{{Coppin}
  et~al.}{2007}]{Coppin+07}
{Coppin} K.~E.~K.,  et~al., 2007, \mn@doi [\apj] {10.1086/519789}, \href
  {https://ui.adsabs.harvard.edu/abs/2007ApJ...665..936C} {665, 936}

\bibitem[\protect\citeauthoryear{{Dobos}, {Csabai}, {Yip}, {Budav{\'a}ri},
  {Wild}  \& {Szalay}}{{Dobos} et~al.}{2012}]{Dobos+12}
{Dobos} L.,  {Csabai} I.,  {Yip} C.-W.,  {Budav{\'a}ri} T.,  {Wild} V.,
  {Szalay} A.~S.,  2012, \mn@doi [\mnras] {10.1111/j.1365-2966.2011.20109.x},
  \href {http://adsabs.harvard.edu/abs/2012MNRAS.420.1217D} {420, 1217}

\bibitem[\protect\citeauthoryear{{Ebbels}, {Le Borgne}, {Pello}, {Ellis},
  {Kneib}, {Smail}  \& {Sanahuja}}{{Ebbels} et~al.}{1996}]{Ebbels+96}
{Ebbels} T.~M.~D.,  {Le Borgne} J.-F.,  {Pello} R.,  {Ellis} R.~S.,  {Kneib}
  J.-P.,  {Smail} I.,   {Sanahuja} B.,  1996, \mn@doi [\mnras]
  {10.1093/mnras/281.4.L75}, \href
  {http://adsabs.harvard.edu/abs/1996MNRAS.281L..75E} {281, L75}

\bibitem[\protect\citeauthoryear{{Faure} et~al.,}{{Faure}
  et~al.}{2009}]{Faure+09}
{Faure} C.,  et~al., 2009, \mn@doi [\apj] {10.1088/0004-637X/695/2/1233}, \href
  {https://ui.adsabs.harvard.edu/abs/2009ApJ...695.1233F} {695, 1233}

\bibitem[\protect\citeauthoryear{{Foreman-Mackey}, {Hogg}, {Lang}  \&
  {Goodman}}{{Foreman-Mackey} et~al.}{2013}]{Foreman+13}
{Foreman-Mackey} D.,  {Hogg} D.~W.,  {Lang} D.,   {Goodman} J.,  2013, \mn@doi
  [\pasp] {10.1086/670067}, \href
  {http://adsabs.harvard.edu/abs/2013PASP..125..306F} {125, 306}

\bibitem[\protect\citeauthoryear{{Frye} et~al.,}{{Frye} et~al.}{2007}]{Frye+07}
{Frye} B.~L.,  et~al., 2007, \mn@doi [\apj] {10.1086/519244}, \href
  {http://adsabs.harvard.edu/abs/2007ApJ...665..921F} {665, 921}

\bibitem[\protect\citeauthoryear{{Furusawa} et~al.,}{{Furusawa}
  et~al.}{2018}]{Furusawa+18}
{Furusawa} H.,  et~al., 2018, \mn@doi [\pasj] {10.1093/pasj/psx079}, \href
  {https://ui.adsabs.harvard.edu/abs/2018PASJ...70S...3F} {70, S3}

\bibitem[\protect\citeauthoryear{{Gemini Observatory} \& {AURA}}{{Gemini
  Observatory} \& {AURA}}{2016}]{giraf+16}
{Gemini Observatory} {AURA} 2016, {Gemini IRAF: Data reduction software for the
  Gemini telescopes} (\mn@eprint {ascl} {1608.006})

\bibitem[\protect\citeauthoryear{{Hezaveh}, {Marrone}  \& {Holder}}{{Hezaveh}
  et~al.}{2012}]{Hezaveh+12}
{Hezaveh} Y.~D.,  {Marrone} D.~P.,   {Holder} G.~P.,  2012, \mn@doi [\apj]
  {10.1088/0004-637X/761/1/20}, \href
  {https://ui.adsabs.harvard.edu/abs/2012ApJ...761...20H} {761, 20}

\bibitem[\protect\citeauthoryear{{Ilbert} et~al.,}{{Ilbert}
  et~al.}{2006}]{Ilbert+06}
{Ilbert} O.,  et~al., 2006, \mn@doi [\aap] {10.1051/0004-6361:20065138}, \href
  {http://adsabs.harvard.edu/abs/2006A%26A...457..841I} {457, 841}

\bibitem[\protect\citeauthoryear{{Ivezi{\'c}} et~al.,}{{Ivezi{\'c}}
  et~al.}{2008}]{Ivezic+08}
{Ivezi{\'c}} Z.,  et~al., 2008, \mn@doi [Serbian Astronomical Journal]
  {10.2298/SAJ0876001I}, \href
  {http://adsabs.harvard.edu/abs/2008SerAJ.176....1I} {176, 1}

\bibitem[\protect\citeauthoryear{{Ivezi{\'c}} et~al.,}{{Ivezi{\'c}}
  et~al.}{2019}]{Ivezic+19}
{Ivezi{\'c}} {\v Z}.,  et~al., 2019, \mn@doi [\apj] {10.3847/1538-4357/ab042c},
  \href {https://ui.adsabs.harvard.edu/abs/2019ApJ...873..111I} {873, 111}

\bibitem[\protect\citeauthoryear{{Jaelani} et~al.,}{{Jaelani}
  et~al.}{2020}]{Jaelani+20}
{Jaelani} A.~T.,  et~al., 2020, arXiv e-prints, \href
  {https://ui.adsabs.harvard.edu/abs/2020arXiv200201611J} {p. arXiv:2002.01611}

\bibitem[\protect\citeauthoryear{{Juri{\'c}} et~al.,}{{Juri{\'c}}
  et~al.}{2017}]{Juric+17}
{Juri{\'c}} M.,  et~al., 2017, in {Lorente} N.~P.~F.,  {Shortridge} K.,
  {Wayth} R.,  eds,  Astronomical Society of the Pacific Conference Series Vol.
  512, Astronomical Data Analysis Software and Systems XXV. p.~279 (\mn@eprint
  {arXiv} {1512.07914})

\bibitem[\protect\citeauthoryear{{Kawanomoto} et~al.,}{{Kawanomoto}
  et~al.}{2018}]{Kawanomoto+18}
{Kawanomoto} S.,  et~al., 2018, \mn@doi [\pasj] {10.1093/pasj/psy056}, \href
  {https://ui.adsabs.harvard.edu/abs/2018PASJ...70...66K} {70, 66}

\bibitem[\protect\citeauthoryear{{Kennicutt}}{{Kennicutt}}{1998}]{Kennicutt+98}
{Kennicutt} Jr. R.~C.,  1998, \mn@doi [\araa] {10.1146/annurev.astro.36.1.189},
  \href {http://adsabs.harvard.edu/abs/1998ARA%26A..36..189K} {36, 189}

\bibitem[\protect\citeauthoryear{{Koester}, {Gladders}, {Hennawi}, {Sharon},
  {Wuyts}, {Rigby}, {Bayliss}  \& {Dahle}}{{Koester} et~al.}{2010}]{Koester+10}
{Koester} B.~P.,  {Gladders} M.~D.,  {Hennawi} J.~F.,  {Sharon} K.,  {Wuyts}
  E.,  {Rigby} J.~R.,  {Bayliss} M.~B.,   {Dahle} H.,  2010, \mn@doi [\apjl]
  {10.1088/2041-8205/723/1/L73}, \href
  {http://adsabs.harvard.edu/abs/2010ApJ...723L..73K} {723, L73}

\bibitem[\protect\citeauthoryear{{Komiyama} et~al.,}{{Komiyama}
  et~al.}{2018}]{Komiyama+18}
{Komiyama} Y.,  et~al., 2018, \mn@doi [\pasj] {10.1093/pasj/psx069}, \href
  {https://ui.adsabs.harvard.edu/abs/2018PASJ...70S...2K} {70, S2}

\bibitem[\protect\citeauthoryear{{Kostrzewa-Rutkowska}, {Wyrzykowski}, {Auger},
  {Collett}  \& {Belokurov}}{{Kostrzewa-Rutkowska} et~al.}{2014}]{Kostrzewa+14}
{Kostrzewa-Rutkowska} Z.,  {Wyrzykowski} {\L}.,  {Auger} M.~W.,  {Collett}
  T.~E.,   {Belokurov} V.,  2014, \mn@doi [\mnras] {10.1093/mnras/stu783},
  \href {https://ui.adsabs.harvard.edu/abs/2014MNRAS.441.3238K} {441, 3238}

\bibitem[\protect\citeauthoryear{{Kuijken} et~al.,}{{Kuijken}
  et~al.}{2019}]{Kuijken+19}
{Kuijken} K.,  et~al., 2019, \mn@doi [\aap] {10.1051/0004-6361/201834918},
  \href {https://ui.adsabs.harvard.edu/abs/2019A&A...625A...2K} {625, A2}

\bibitem[\protect\citeauthoryear{{Marques-Chaves} et~al.,}{{Marques-Chaves}
  et~al.}{2017}]{Marques+17}
{Marques-Chaves} R.,  et~al., 2017, \mn@doi [\apjl]
  {10.3847/2041-8213/834/2/L18}, \href
  {https://ui.adsabs.harvard.edu/abs/2017ApJ...834L..18M} {834, L18}

\bibitem[\protect\citeauthoryear{{Mason}, {C{\^o}t{\'e}}, {Kissler-Patig},
  {Levenson}, {Adamson}, {Emmanuel}  \& {Crabtree}}{{Mason}
  et~al.}{2014}]{Mason+14}
{Mason} R.~E.,  {C{\^o}t{\'e}} S.,  {Kissler-Patig} M.,  {Levenson} N.~A.,
  {Adamson} A.,  {Emmanuel} C.,   {Crabtree} D.,  2014, in Observatory
  Operations: Strategies, Processes, and Systems V. p. 914910 (\mn@eprint
  {arXiv} {1408.5916}), \mn@doi{10.1117/12.2057145}

\bibitem[\protect\citeauthoryear{{Mehlert} et~al.,}{{Mehlert}
  et~al.}{2001}]{Mehlert+01}
{Mehlert} D.,  et~al., 2001, \mn@doi [\aap] {10.1051/0004-6361:20011286}, \href
  {http://adsabs.harvard.edu/abs/2001A%26A...379...96M} {379, 96}

\bibitem[\protect\citeauthoryear{{Miyazaki} et~al.,}{{Miyazaki}
  et~al.}{2018}]{Miyazaki+18}
{Miyazaki} S.,  et~al., 2018, \mn@doi [\pasj] {10.1093/pasj/psx063}, \href
  {http://adsabs.harvard.edu/abs/2018PASJ...70S...1M} {70, S1}

\bibitem[\protect\citeauthoryear{{More}, {Jahnke}, {More}, {Gallazzi}, {Bell},
  {Barden}  \& {H{\"a}u{\ss}ler}}{{More} et~al.}{2011}]{More+11}
{More} A.,  {Jahnke} K.,  {More} S.,  {Gallazzi} A.,  {Bell} E.~F.,  {Barden}
  M.,   {H{\"a}u{\ss}ler} B.,  2011, \mn@doi [\apj]
  {10.1088/0004-637X/734/1/69}, \href
  {https://ui.adsabs.harvard.edu/abs/2011ApJ...734...69M} {734, 69}

\bibitem[\protect\citeauthoryear{{More} et~al.,}{{More} et~al.}{2017}]{More+17}
{More} A.,  et~al., 2017, \mn@doi [\mnras] {10.1093/mnras/stw2924}, \href
  {http://adsabs.harvard.edu/abs/2017MNRAS.465.2411M} {465, 2411}

\bibitem[\protect\citeauthoryear{{Oguri}}{{Oguri}}{2010}]{Oguri+10}
{Oguri} M.,  2010, \mn@doi [\pasj] {10.1093/pasj/62.4.1017}, \href
  {http://adsabs.harvard.edu/abs/2010PASJ...62.1017O} {62, 1017}

\bibitem[\protect\citeauthoryear{{Oguri} et~al.,}{{Oguri}
  et~al.}{2018}]{Oguri+18}
{Oguri} M.,  et~al., 2018, \mn@doi [\pasj] {10.1093/pasj/psx042}, \href
  {https://ui.adsabs.harvard.edu/abs/2018PASJ...70S..20O} {70, S20}

\bibitem[\protect\citeauthoryear{{Oldham} et~al.,}{{Oldham}
  et~al.}{2017}]{Oldham+17}
{Oldham} L.,  et~al., 2017, \mn@doi [\mnras] {10.1093/mnras/stw2832}, \href
  {https://ui.adsabs.harvard.edu/abs/2017MNRAS.465.3185O} {465, 3185}

\bibitem[\protect\citeauthoryear{{Ono} et~al.,}{{Ono} et~al.}{2018}]{Ono+18}
{Ono} Y.,  et~al., 2018, \mn@doi [\pasj] {10.1093/pasj/psx103}, \href
  {https://ui.adsabs.harvard.edu/abs/2018PASJ...70S..10O} {70, S10}

\bibitem[\protect\citeauthoryear{{Peng}, {Ho}, {Impey}  \& {Rix}}{{Peng}
  et~al.}{2002}]{Peng+02}
{Peng} C.~Y.,  {Ho} L.~C.,  {Impey} C.~D.,   {Rix} H.-W.,  2002, \mn@doi [\aj]
  {10.1086/340952}, \href {http://adsabs.harvard.edu/abs/2002AJ....124..266P}
  {124, 266}

\bibitem[\protect\citeauthoryear{{Pettini}, {Steidel}, {Adelberger},
  {Dickinson}  \& {Giavalisco}}{{Pettini} et~al.}{2000}]{Pettini+00}
{Pettini} M.,  {Steidel} C.~C.,  {Adelberger} K.~L.,  {Dickinson} M.,
  {Giavalisco} M.,  2000, \mn@doi [\apj] {10.1086/308176}, \href
  {http://adsabs.harvard.edu/abs/2000ApJ...528...96P} {528, 96}

\bibitem[\protect\citeauthoryear{{Pettini}, {Rix}, {Steidel}, {Adelberger},
  {Hunt}  \& {Shapley}}{{Pettini} et~al.}{2002}]{Pettini+02}
{Pettini} M.,  {Rix} S.~A.,  {Steidel} C.~C.,  {Adelberger} K.~L.,  {Hunt}
  M.~P.,   {Shapley} A.~E.,  2002, \mn@doi [\apj] {10.1086/339355}, \href
  {http://adsabs.harvard.edu/abs/2002ApJ...569..742P} {569, 742}

\bibitem[\protect\citeauthoryear{{Quider}, {Pettini}, {Shapley}  \&
  {Steidel}}{{Quider} et~al.}{2009}]{Quider+09}
{Quider} A.~M.,  {Pettini} M.,  {Shapley} A.~E.,   {Steidel} C.~C.,  2009,
  \mn@doi [\mnras] {10.1111/j.1365-2966.2009.15234.x}, \href
  {https://ui.adsabs.harvard.edu/abs/2009MNRAS.398.1263Q} {398, 1263}

\bibitem[\protect\citeauthoryear{{Quider}, {Shapley}, {Pettini}, {Steidel}  \&
  {Stark}}{{Quider} et~al.}{2010}]{Quider+10}
{Quider} A.~M.,  {Shapley} A.~E.,  {Pettini} M.,  {Steidel} C.~C.,   {Stark}
  D.~P.,  2010, \mn@doi [\mnras] {10.1111/j.1365-2966.2009.16005.x}, \href
  {https://ui.adsabs.harvard.edu/abs/2010MNRAS.402.1467Q} {402, 1467}

\bibitem[\protect\citeauthoryear{{Reddy} \& {Steidel}}{{Reddy} \&
  {Steidel}}{2009}]{Reddy+09}
{Reddy} N.~A.,  {Steidel} C.~C.,  2009, \mn@doi [\apj]
  {10.1088/0004-637X/692/1/778}, \href
  {http://adsabs.harvard.edu/abs/2009ApJ...692..778R} {692, 778}

\bibitem[\protect\citeauthoryear{{Richard}, {Stark}, {Ellis}, {George},
  {Egami}, {Kneib}  \& {Smith}}{{Richard} et~al.}{2008}]{Richard+08}
{Richard} J.,  {Stark} D.~P.,  {Ellis} R.~S.,  {George} M.~R.,  {Egami} E.,
  {Kneib} J.-P.,   {Smith} G.~P.,  2008, \mn@doi [\apj] {10.1086/591312}, \href
  {https://ui.adsabs.harvard.edu/abs/2008ApJ...685..705R} {685, 705}

\bibitem[\protect\citeauthoryear{{Richard}, {Jones}, {Ellis}, {Stark},
  {Livermore}  \& {Swinbank}}{{Richard} et~al.}{2011}]{Richard+11}
{Richard} J.,  {Jones} T.,  {Ellis} R.,  {Stark} D.~P.,  {Livermore} R.,
  {Swinbank} M.,  2011, \mn@doi [\mnras] {10.1111/j.1365-2966.2010.18161.x},
  \href {https://ui.adsabs.harvard.edu/abs/2011MNRAS.413..643R} {413, 643}

\bibitem[\protect\citeauthoryear{{S{\'a}nchez-Bl{\'a}zquez}
  et~al.,}{{S{\'a}nchez-Bl{\'a}zquez} et~al.}{2006}]{Sanchez+06}
{S{\'a}nchez-Bl{\'a}zquez} P.,  et~al., 2006, \mn@doi [\mnras]
  {10.1111/j.1365-2966.2006.10699.x}, \href
  {http://adsabs.harvard.edu/abs/2006MNRAS.371..703S} {371, 703}

\bibitem[\protect\citeauthoryear{{Sawicki} \& {Thompson}}{{Sawicki} \&
  {Thompson}}{2006}]{Sawicki+06}
{Sawicki} M.,  {Thompson} D.,  2006, \mn@doi [\apj] {10.1086/500999}, \href
  {https://ui.adsabs.harvard.edu/abs/2006ApJ...642..653S} {642, 653}

\bibitem[\protect\citeauthoryear{{Seitz}, {Saglia}, {Bender}, {Hopp}, {Belloni}
   \& {Ziegler}}{{Seitz} et~al.}{1998}]{Seitz+98}
{Seitz} S.,  {Saglia} R.~P.,  {Bender} R.,  {Hopp} U.,  {Belloni} P.,
  {Ziegler} B.,  1998, \mn@doi [\mnras] {10.1046/j.1365-8711.1998.01443.x},
  \href {https://ui.adsabs.harvard.edu/abs/1998MNRAS.298..945S} {298, 945}

\bibitem[\protect\citeauthoryear{{Shapley}, {Steidel}, {Pettini}  \&
  {Adelberger}}{{Shapley} et~al.}{2003}]{Shapley+03}
{Shapley} A.~E.,  {Steidel} C.~C.,  {Pettini} M.,   {Adelberger} K.~L.,  2003,
  \mn@doi [\apj] {10.1086/373922}, \href
  {http://adsabs.harvard.edu/abs/2003ApJ...588...65S} {588, 65}

\bibitem[\protect\citeauthoryear{{Shibuya}, {Ouchi}  \& {Harikane}}{{Shibuya}
  et~al.}{2015}]{Shibuya+15}
{Shibuya} T.,  {Ouchi} M.,   {Harikane} Y.,  2015, \mn@doi [\apjs]
  {10.1088/0067-0049/219/2/15}, \href
  {http://adsabs.harvard.edu/abs/2015ApJS..219...15S} {219, 15}

\bibitem[\protect\citeauthoryear{{Siana} et~al.,}{{Siana}
  et~al.}{2009}]{Siana+09}
{Siana} B.,  et~al., 2009, \mn@doi [\apj] {10.1088/0004-637X/698/2/1273}, \href
  {https://ui.adsabs.harvard.edu/abs/2009ApJ...698.1273S} {698, 1273}

\bibitem[\protect\citeauthoryear{{Smail} et~al.,}{{Smail}
  et~al.}{2007}]{Smail+07}
{Smail} I.,  et~al., 2007, \mn@doi [\apjl] {10.1086/510902}, \href
  {http://adsabs.harvard.edu/abs/2007ApJ...654L..33S} {654, L33}

\bibitem[\protect\citeauthoryear{{Sonnenfeld}, {Gavazzi}, {Suyu}, {Treu}  \&
  {Marshall}}{{Sonnenfeld} et~al.}{2013a}]{Sonnenfeld+13a}
{Sonnenfeld} A.,  {Gavazzi} R.,  {Suyu} S.~H.,  {Treu} T.,   {Marshall} P.~J.,
  2013a, \mn@doi [\apj] {10.1088/0004-637X/777/2/97}, \href
  {https://ui.adsabs.harvard.edu/abs/2013ApJ...777...97S} {777, 97}

\bibitem[\protect\citeauthoryear{{Sonnenfeld}, {Treu}, {Gavazzi}, {Suyu},
  {Marshall}, {Auger}  \& {Nipoti}}{{Sonnenfeld}
  et~al.}{2013b}]{Sonnenfeld+13b}
{Sonnenfeld} A.,  {Treu} T.,  {Gavazzi} R.,  {Suyu} S.~H.,  {Marshall} P.~J.,
  {Auger} M.~W.,   {Nipoti} C.,  2013b, \mn@doi [\apj]
  {10.1088/0004-637X/777/2/98}, \href
  {https://ui.adsabs.harvard.edu/abs/2013ApJ...777...98S} {777, 98}

\bibitem[\protect\citeauthoryear{{Sonnenfeld} et~al.,}{{Sonnenfeld}
  et~al.}{2018}]{Sonnenfeld+18}
{Sonnenfeld} A.,  et~al., 2018, \mn@doi [\pasj] {10.1093/pasj/psx062}, \href
  {http://adsabs.harvard.edu/abs/2018PASJ...70S..29S} {70, S29}

\bibitem[\protect\citeauthoryear{{Sonnenfeld}, {Jaelani}, {Chan}, {More},
  {Suyu}, {Wong}, {Oguri}  \& {Lee}}{{Sonnenfeld} et~al.}{2019}]{Sonnenfeld+19}
{Sonnenfeld} A.,  {Jaelani} A.~T.,  {Chan} J.,  {More} A.,  {Suyu} S.~H.,
  {Wong} K.~C.,  {Oguri} M.,   {Lee} C.-H.,  2019, \mn@doi [\aap]
  {10.1051/0004-6361/201935743}, \href
  {https://ui.adsabs.harvard.edu/abs/2019A&A...630A..71S} {630, A71}

\bibitem[\protect\citeauthoryear{{Stark}, {Ellis}, {Richard}, {Kneib}, {Smith}
  \& {Santos}}{{Stark} et~al.}{2007}]{Stark+07}
{Stark} D.~P.,  {Ellis} R.~S.,  {Richard} J.,  {Kneib} J.-P.,  {Smith} G.~P.,
  {Santos} M.~R.,  2007, \mn@doi [\apj] {10.1086/518098}, \href
  {https://ui.adsabs.harvard.edu/abs/2007ApJ...663...10S} {663, 10}

\bibitem[\protect\citeauthoryear{{Steidel}, {Adelberger}, {Dickinson},
  {Giavalisco}, {Pettini}  \& {Kellogg}}{{Steidel} et~al.}{1998}]{Steidel+98}
{Steidel} C.~C.,  {Adelberger} K.~L.,  {Dickinson} M.,  {Giavalisco} M.,
  {Pettini} M.,   {Kellogg} M.,  1998, \mn@doi [\apj] {10.1086/305073}, \href
  {http://adsabs.harvard.edu/abs/1998ApJ...492..428S} {492, 428}

\bibitem[\protect\citeauthoryear{{Steidel}, {Shapley}, {Pettini}, {Adelberger},
  {Erb}, {Reddy}  \& {Hunt}}{{Steidel} et~al.}{2004}]{Steidel+04}
{Steidel} C.~C.,  {Shapley} A.~E.,  {Pettini} M.,  {Adelberger} K.~L.,  {Erb}
  D.~K.,  {Reddy} N.~A.,   {Hunt} M.~P.,  2004, \mn@doi [\apj]
  {10.1086/381960}, \href {http://adsabs.harvard.edu/abs/2004ApJ...604..534S}
  {604, 534}

\bibitem[\protect\citeauthoryear{{Tody}}{{Tody}}{1986}]{Tody+86}
{Tody} D.,  1986, in {Crawford} D.~L.,  ed.,  \procspie Vol. 627,
  Instrumentation in astronomy VI. p.~733, \mn@doi{10.1117/12.968154}

\bibitem[\protect\citeauthoryear{{Tody}}{{Tody}}{1993}]{Tody+93}
{Tody} D.,  1993, in {Hanisch} R.~J.,  {Brissenden} R.~J.~V.,   {Barnes} J.,
  eds,  Astronomical Society of the Pacific Conference Series Vol. 52,
  Astronomical Data Analysis Software and Systems II. p.~173

\bibitem[\protect\citeauthoryear{{Wolf}, {Martinez}, {Bullock}, {Kaplinghat},
  {Geha}, {Mu{\~n}oz}, {Simon}  \& {Avedo}}{{Wolf} et~al.}{2010}]{Wolf+10}
{Wolf} J.,  {Martinez} G.~D.,  {Bullock} J.~S.,  {Kaplinghat} M.,  {Geha} M.,
  {Mu{\~n}oz} R.~R.,  {Simon} J.~D.,   {Avedo} F.~F.,  2010, \mn@doi [\mnras]
  {10.1111/j.1365-2966.2010.16753.x}, \href
  {https://ui.adsabs.harvard.edu/abs/2010MNRAS.406.1220W} {406, 1220}

\bibitem[\protect\citeauthoryear{{Wong} et~al.,}{{Wong} et~al.}{2018}]{Wong+18}
{Wong} K.~C.,  et~al., 2018, \mn@doi [\apj] {10.3847/1538-4357/aae381}, \href
  {http://adsabs.harvard.edu/abs/2018ApJ...867..107W} {867, 107}

\bibitem[\protect\citeauthoryear{{Yee}, {Ellingson}  \& {Carlberg}}{{Yee}
  et~al.}{1996}]{Yee+96}
{Yee} H.~K.~C.,  {Ellingson} E.,   {Carlberg} R.~G.,  1996, \mn@doi [\apjs]
  {10.1086/192259}, \href
  {https://ui.adsabs.harvard.edu/abs/1996ApJS..102..269Y} {102, 269}

\makeatother
\end{thebibliography}

\appendix
\section{Expected lensed (compact) LBG detections in the HSC Survey}\label{app:expected}
In order to estimate the expected number of unusually compact lensed sources to be found in the HSC Survey, we used the number of LBGs and typical optical depths for lensing following Equation 1 from \cite{More+11}. The optical depth was taken from Table 2 of \cite{Faure+09} (for $r_{\rm arc}<1.5$~arcsec, $z_s=2$, intermediate convergence region) and the number density of sources is calculated from the LBG luminosity function at $z\sim4$ from \cite{Ono+18}. Given the HSC Survey design, we take sources with apparent magnitude brighter than $m_{i, \rm s}\sim 26.4$ within the full area of 1,400 deg$^2$. Note, that this corresponds to an absolute magnitude of $-19.6$ in \fref{fig:shibuya15}. Next, we restrict ourselves to effective radii around $r_{e, \rm s}=0.1-0.2$ kpc by using the PDF given in Figure 6 of S15. The resulting number of unusually compact lensed systems in the HSC Survey is $\gtrsim 40$. In reality, sources fainter than the limiting magnitude can still end up in the sample which will increase this number further. We also note that the number estimate is also very sensitive to the choice of optical depth for lensing and should be taken with a grain of salt.

\section{Corner plot of the lens model}\label{app:sugohig}

\begin{figure*}
\begin{center}
\includegraphics[width=\textwidth]{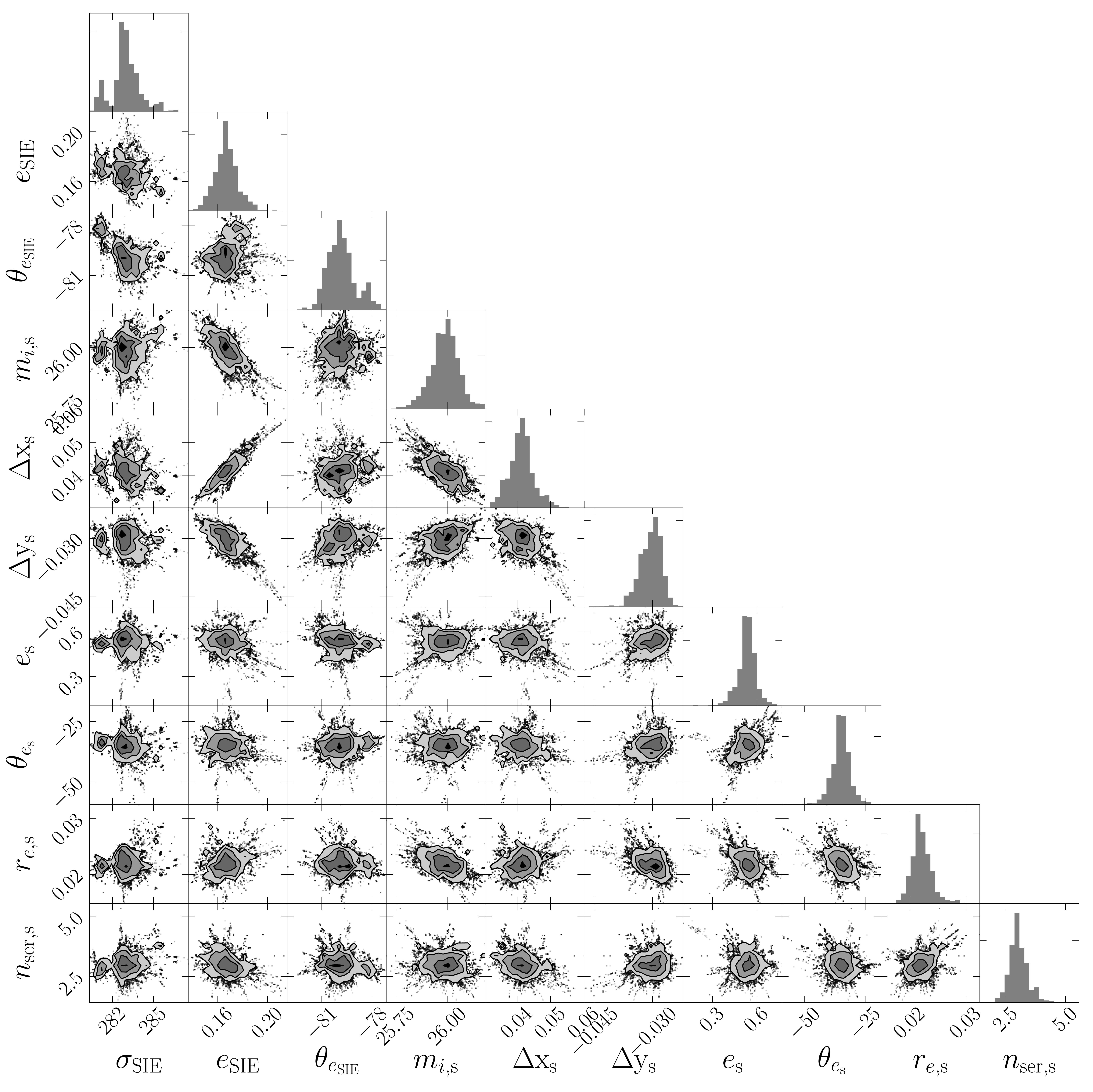}\\
\caption{\label{fig:chaincorner} Corner plot of the one- and two-dimensional probability density distributions for the SIE model parameters from lens model fitting results in the $i$-band with {\sc glafic}. The contours represent the 1$\sigma$, 2$\sigma$ and 3$\sigma$. The parameters (from left to right) are lens velocity dispersion in units of km s$^{-1}$, lens ellipticity, lens position angle, source magnitude, source centroid relative to the lens centroid, source ellipticity, source position angle, source effective radius in units of arcsec, and source S\'{e}rsic index.}
\end{center}
\end{figure*}


\bsp	
\label{lastpage}
\end{document}